# Loss Compensation and Super-Resolution with Excitations at Complex Frequencies

Seunghwi Kim[1†], Yu-Gui Peng[1,2†], Simon Yves[1], and Andrea Alù[1,3*]

[1]*Photonics Initiative, Advanced Science Research Center, City University of New York, New York, New York 10031, USA*

[2] *School of Physics, Huazhong University of Science and Technology, 430074, Wuhan, China*

[3]*Physics Program, Graduate Center, City University of New York, New York, NY 10016, USA*

[†]These authors contributed equally to this work

[*]Corresponding author. Email: aalu@gc.cuny.edu

**Abstract:** Abbe diffraction limit fundamentally bounds the resolution of conventional optical imaging and spectroscopic systems. Along the years, several schemes have been introduced to overcome this limit, each offering opportunities and trade-offs. Metamaterials have been proposed as a promising platform in this context, in principle enabling superlenses capable of drastic resolution enhancements. However, their performance is severely hindered by material loss and nonlocal phenomena, which become increasingly more detrimental as we attempt to image more subwavelength details. Active metamaterials have been explored to overcome these challenges, however material gain introduces other obstacles, e.g., instabilities, nonlinearities and noise. Here, we demonstrate that the temporal excitation of passive superlenses using signals oscillating at complex frequencies compensates material loss, leading to resolution enhancement. Our results demonstrate that virtual gain stemming from tailored forms of excitation can tackle the impact of loss in superlenses. More broadly, our work opens promising avenues for loss compensation in metamaterials, with a broad range of applications extendable to optics and nanophotonic systems.

***Introduction*** - The excitement around metamaterials has been fueled by the discovery that their unusual electromagnetic properties may in principle realize superlenses with no geometrical aberrations and unlimited resolution, holding the promise to revolutionize imaging and spectroscopy [1-4]. Conventional lenses focus only propagating waves, resulting in the rejection of image details associated with transverse wavenumbers larger than $k_0 = \omega/c$, where $\omega$ is the angular frequency of the impinging waves and $c$ is the velocity of light in the background medium. These waves are evanescent, they decay away from the object and cannot be focused. Hence, the minimum resolvable feature through a conventional lens is comparable to the wavelength of operation $\Delta_{\min} = 2\pi/k_0 \approx \lambda_{op}$. Quite remarkably, a planar metamaterial slab with a negative index of refraction has been shown to focus propagating waves as a regular lens [5], but also to restore the amplitude of evanescent waves impinging on it, implying that such a structure can in principle achieve complete restoration of the image details at its focal plane, as sketched in Fig. 1A [1]. In this scenario, the minimum resolvable feature size $\Delta_{\min}$ becomes deeply subwavelength, not complying with the conventional resolution limit [1,6]. This intriguing phenomenon is rooted into the exotic wave-matter interactions enabled by negative refractive index materials [5], which can be implemented as metamaterials in various frequency ranges and within several material platforms [7-10]. While a negative index of refraction requires the implementation of negative electric and magnetic permittivity at the same frequency, it has also been shown that a *poor-man superlens* can be achieved for one polarization of light using negative-permittivity thin films, in which the amplification of evanescent waves is supported by surface polariton resonances at the back of the film [3,4].

Despite these exciting prospects, successful implementations of metamaterial superlenses have been limited to date, and their application in practical imaging setups has been lagging. The main reason for this lack of development resides in the fact that the resolution of superlenses is hindered in practice by material loss and nonlocalities [11-13]. In passive metamaterials, the presence of loss is inevitable due to material absorption and scattering losses from inhomogeneities, defects, dislocations and surface roughness, and larger transverse wave numbers are increasingly more sensitive to a given level of loss [12]. Material nonlocalities are also inevitable, due to the granularity of the constituent lattice of the engineered material: evanescent waves carrying large transverse wave numbers are associated with rapidly varying field variations that do not capture the effective properties of the metamaterial once they become comparable with the lattice constant [13]. Hence, material loss and nonlocality fundamentally limit the maximum accessible resolution of metamaterial superlenses [12,14]. Figure 1B schematizes this concept, showing that evanescent fields can only be partially amplified inside such an imperfect superlens, limiting the resolution at the image plane.

Anisotropic multilayers consisting of thin dielectrics alternated with negative-index materials have been proposed as a way to partially tackle this issue, yet their resolution is still fundamentally limited [15,16]. Since the impact of material loss is typically stronger than the one of nonlocality, a way to enhance the resolution may be the use of material gain as a way to compensate for the loss [17,18]. Yet, gain is scarce and difficult to control, and in some instances the introduction of gain may even affect the negative refractive index properties of the underlying media [19,20]. Also, instabilities, additional noise and clamping of population inversion are hardly compatible with practical imaging systems [21].

In the past few years, it has been noticed that it may be possible to go around some of these limitations by controlling the evolution in time of the incident signals. For instance, time-reversal techniques have been used to mimic the growth of the evanescent portion of the spectrum, exceeding the diffraction limit [22,23]. These techniques, however, rely on active systems, which imply a processing delay of the received signals and the need for complex and energy-intensive techniques for time-reversal. Transmission matrix and adaptive optics using spatial-light modulators have also enabled imaging beyond the diffraction limit, but the complete characterization of the scattering medium is a prerequisite. Superoscillations have also been explored to enhance the imaging resolution by exciting multiple sub-diffraction modes [24,25]. This approach, however, can target specific details of an object, with limitations in the case of complex images. It has also been theoretically shown that, when the monochromatic excitation of a negative-index slab is abruptly shut down [26] or exponentially tapered [27], the resolution at the image plane can be enhanced.

Here, we demonstrate that it is possible to compensate for the detrimental effect of material loss on superlenses by introducing *virtual gain*, associated with tailored excitations oscillating at complex frequencies. In a different context, we have recently shown that it is possible to overcome the limitations of passive scattering systems by exciting them with complex frequency signals [28,29]. By analytically continuing the response function of a linear system, it is possible to unveil singularities in the complex frequency plane, which can be engaged with signals oscillating periodically with frequency $\text{Re}[\omega]$ while their amplitude is modulated by a growing or decaying exponential factor $\text{Im}[\omega]$. Provided that the system is tailored to reach *quasi-steady state* at this complex frequency, i.e., that after a transient the system converges to a response oscillating at the same complex frequency as the input, such excitations can provide an effect analogous to material gain (loss) for $\text{Im}[\omega] < 0$ ( $\text{Im}[\omega] > 0$ ) [30,31]. This principle can support coherent perfect absorption without material loss [28,32], *virtual* PT-symmetry without material gain [30], *virtual* critical coupling [33], negative radiation pressure [34], and exotic scattering features beyond the limits of passive objects [31]. Here, we demonstrate that excitations at complex frequencies can be used to compensate material loss in metamaterials and demonstrate that such *virtual gain* can be exploited to restore the deeply subwavelength resolution of superlenses, as schematically shown in Fig. 1C. Quite importantly, this effect does not lead to instabilities or noise as in the case of material gain, being based on linear, passive, time-invariant systems.

***Results*** – In order to shed light on the phenomenon, we start by considering a poor-man superlens formed by a non-magnetic planar thin slab of thickness $h = 10\,\text{nm}$ with Drude permittivity $\varepsilon(\omega) = \varepsilon_h - (\varepsilon_s - \varepsilon_h)\omega_p^2 / (\omega^2 + i2\omega\alpha)$, where $\varepsilon_h = 5.45$, $\varepsilon_h = 6.18$, $\omega_p$ is the plasma frequency, and $\alpha = 5\times 10^{14}\,\text{rad/s}$ is the loss rate. In order to study its imaging properties, we consider the transmission from object to image plane as a function of the transverse wavenumber $k_x$ [1]

$$T(k_x, \omega) = \frac{4 k_z q_z \varepsilon \exp(iq_z h)}{(q_z + \varepsilon k_z)^2 - (q_z - \varepsilon k_z)^2 \exp(2iq_z h)}, \tag{1}$$

where $q_z = \sqrt{\mu\varepsilon k_0^2 - k_x^2}$ and $k_z = \sqrt{k_0^2 - k_x^2}$ are the wavenumbers inside and outside the medium in the longitudinal direction $z$. In the absence of material loss ($\alpha = 0$), the evanescent waves impinging on the slab with $k_x > k_0$ are amplified in the $z$-direction, resulting in compensation of their decay in air, and ensuring that their amplitude is restored on the image plane. In the presence of loss, however, the evanescent fields are less amplified, consistent with Fig. 1B and the red curve in Fig. 2A, and a cut-off emerges for large wave numbers $k_{cutoff} = -\ln\left[\left|\text{Im}(\varepsilon)/2\right|^2\right]/(2h) = 3.1 k_o$ [12,14], limiting the device resolution.

Next, we explore an excitation at complex frequency $E_i(t) = E_o \exp(-i \operatorname{Re}[\omega_c]t) \exp(\operatorname{Im}[\omega_c]t)$, whose negative imaginary part describes an exponentially decaying envelope $\gamma = -\operatorname{Im}[\omega_c]$. Assuming that we can analytically continue Eq. (1), Figure 2A shows the calculated transmission coefficient $\tilde{T}(k_x, \omega) = T(k_x, \omega)/T_{lossless}$, normalized to the lossless scenario $T_{lossless}$, for different values of $\gamma$. For $\gamma = 0$, the transmission has a cut-off around $k_x \approx 10 k_0$ (red curve). However, as we move in the complex frequency plane, the transmission significantly improves, e.g., $\gamma = 0.75\alpha$ and $\gamma = 0.9\alpha$ for blue and green curves in Fig. 2A, respectively. Full recovery of the imaging properties is found when the decay rate $\gamma$ is equal to the material loss rate $\alpha$ (yellow curve), restoring the ideal resolving features of a lossless superlens [27]. These results are meaningful only under the assumption that the superlens output reaches a quasi-steady state that oscillates at the same complex frequency as the excitation, for which the analytical continuation of Eq. (1) has a physical meaning [31]. Indeed, for $\gamma = 0.75\alpha$, $\gamma = 0.9\alpha$, and $\gamma = \alpha$, the Drude permittivity evaluated at such complex frequency is $\varepsilon = -0.991 + i0.277$, $\varepsilon = -0.999 + i0.111$, and $\varepsilon = -1$, respectively, compensating for the material loss and explaining why the superlens operation is restored [1].

This result indicates that material absorption can be in principle compensated by a suitably chosen complex frequency excitation. In Fig. 2B, we show the density plot of normalized transmission in the complex frequency plane for a large transverse wavenumber $k_{x0} = 30 k_o$, finding that the transmission becomes unity for $\omega_c = 5.77 \times 10^{15} - i5 \times 10^{14}$ rad/s. Hence, we can expect that, if the slab reaches quasi-steady state, we may expect a complete restoration of its imaging properties. In this scenario, the material loss is exactly compensated by *virtual gain* stemming from the excitation. Fig. 2C shows the transmission density plot as a function of $k_x$ and $\gamma$, which becomes unity for all wave numbers when $\gamma = \alpha$. Equivalently, Fig. 2D shows the point spread function for the scenarios considered in Fig. 2A, i.e., the normalized amplitude on the image plane for a delta function source as we vary the decay rates $\gamma = 0$ (harmonic), $\gamma = 0.75\alpha$, $\gamma = 0.9\alpha$ and $\gamma = a$. The spread functions get narrower as the decay rate approaches the loss rate, and the width of the spread function is minimal for $\gamma = \alpha$, confirming that the right amount of *virtual* gain enables complete recovery of the performance of lossy superlenses. Further discussion on a realistic scenario using SiC slabs is discussed in Supplementary Section S1.

In order to demonstrate this phenomenon in a practical setup, we explore its implementation in an acoustic platform. We consider a holey acoustic metamaterial with deeply subwavelength periodic square holes and overall thickness $h$, as shown in Fig. 3A. Such a structure has been proposed to realize an acoustic superlens with resolution well beyond the diffraction limit by relying on acoustic spoof plasmons mediating the transfer of the entire spectrum of an image in combination with a longitudinal Fabry-Perot (FP) resonance [35]. The inset of Fig. 3A illustrates the zoomed cross-section displaying the square lattice with periodicity $d$ and air hole size $a$ surrounded by rigid boundaries. The transmission as a function of transverse wavenumber through such holey structures in the subwavelength regime $\lambda \gg a, d$ is given by [36]

$$T_a(k_x, \omega) = \frac{4 Y_{hole} Y_0 \left|S^{(0)}\right|^2 \exp(i q_{z,a} h)}{\left(Y_{hole} + Y_0 \left|S^{(0)}\right|^2\right)^2 - \left(Y_{hole} - Y_0 \left|S^{(0)}\right|^2\right)^2 \exp(2 i g_{z,a} h)}, \qquad (2)$$

Where $Y_0$ and $Y_{hole}$ are the admittance of the $0^{th}$ diffraction order mode and of the holes, respectively, and $S^{(0)} = \sqrt{a/d} \, \text{sinc}(k_x a/2)$ is the overlap integral of the fundamental and incident modes (see Supplementary Sections S2) [35,36]. The expression in Eq. (2) is analogous to the expression in Eq. (1), with similar features in terms of resolution cut-off. One remarkable difference is that the wavenumber inside the metamaterial in the z-direction is $q_{z,a} = k_0$ for all transverse wave numbers, due to spatial dispersion, hence this device does not amplify evanescent fields, but simply makes them propagative. In order to ensure full transmission of all transverse wave numbers, we rely on a longitudinal Fabry-Perot resonance, $q_{z,a} = m\pi/h$ where $m$ is an integer, exploiting the fact that all propagating waves travel with the same phase velocity. Figure 3B (left) shows the transmission versus normalized transverse wavenumber $k_x/k_0$ for the lossless scenario (yellow curve) at the first-order Fabry-Perot resonance for $h$ = 10 cm, $a$ = 0.15 cm and $d$ = 0.2 cm, in the case of harmonic excitation. We compare the ideal imaging response (blue line), which neglects nonlocal phenomena, with the transmission of the lossless metamaterial (yellow line), with a cut-off determined by the periodicity $k_x \leq 2\pi/d$ (red dotted line), past which Eq. (2) does not apply any longer. In the presence of realistic material loss with absorption rate $\alpha$, the dispersion decays faster (orange line), worsening the resolution. On the right panel of Fig. 3B we study the response in the complex frequency plane, for different excitation decay rates $\gamma = 0.5\alpha$, $\gamma = 0.75\alpha$ and $\gamma = \alpha$. Indeed, consistent with the previous results, moving to complex frequencies enhances the response and restores the imaging functionality to the ideal lossless scenario in which nonlocality dominates. In particular, for a decay rate equal to the loss rate $\gamma = \alpha$ (yellow), the transmission becomes identical to the lossless scenario, clearly evidencing the potential of virtual gain also in this acoustic platform.

So far, these results have been assuming that analytically continuing Eq. (2) yields a physically meaningful output. However, complex frequency signals are not bounded, hence they necessarily have to start at a given instant in time and, due to their decay and their non-orthogonality, there is no guarantee that the system will converge within a finite transient to an output oscillating at the same complex frequency [31]. Hence, in order to verify our theoretical predictions, we performed full-wave time-domain simulations for a realistic excitation starting at $t=0$ and studied the temporal evolution of the response. In order to test the imaging features, we test the metamaterial with an object composed of three parallel lines perforated over a thin rectangle solid plate (Fig. 3C). We place this object next to the holey acoustic superlens and excite it with acoustic waves with same carrier frequency $\text{Re}[\omega_c]$, set at 1715 Hz in order to excite the first FP resonance of the slab, and different amplitude modulations to control the imaginary part $\gamma$ of the excitation frequency. The width of the lines is 0.4 cm and the inter-line spacings are 0.4 cm and 0.2 cm. The transmitted acoustic fields are evaluated at the image plane in real-time, and we are particularly interested in the quasi-steady-state regime, for which the output response after a transient oscillates at the same complex frequency as the excitation (see Supplementary Section S3) [30,31]. Material loss is introduced in the simulations through thermos-viscous phenomena, which are expected to dominate the material loss mechanism in practical structures. The output sound intensity image under monochromatic excitation ($\gamma=0$ in Fig. 3D) shows a clear limit in resolution, making it impossible to differentiate the strips of the original object (dotted lines). However, as we increase $\gamma$, the spatial resolution of the transmitted image is dramatically enhanced. In particular, for $\gamma=\alpha$ the transmitted intensity map is clearly comparable to the image obtained in the case of a lossless system with monochromatic excitation (right panel of Fig. 3D), recovering the subwavelength spatial features of the object. This is further evidenced by the image cuts at $x=0$ presented in Fig. 3E, where the intensity profiles of the lossless and virtual gain scenarios (green dotted and yellow curve, respectively) are compared. By computing the 2D spatial Fourier transform (FT) of the intensity maps in Fig. 3D, we extract the transverse spatial Fourier components of each image (Fig. 3F). The FT maps indeed show stronger contributions from the large transverse wave-numbers as the decay rate $\gamma$ approaches the loss rate $\alpha$, providing clear evidence of the resolution enhancement associated with complex frequency excitation in a realistic setup. Notably, the FT map for $\gamma = \alpha$ is almost identical to the lossless scenario in the right panel of Fig. 3F. We further analyze these results in Supplementary Sections S2 and S3. These simulations confirm that analytical continuation of Eq. (2) is indeed meaningful, and the response of the superlens does reach a quasi-steady state in which virtual gain compensates the detrimental effects of material loss.

We have realized a 3D-printed holey-structured acoustic metamaterial, which expectedly has non-negligible loss due to thermos-viscous dissipation as well as fabrication inaccuracies. We placed the three-line object (left top panel in Fig. 4A) on one side of the metamaterial right next to the holes and added a screen sheet on the other side to measure the transmitted pressure fields with a 3D laser vibrometer (see Supplementary Section S4 for detailed information on our experiment). The measured normalized acoustic energy averaged over one cycle for various complex frequency excitations with $\gamma = 0, 30, 60, 90, 120, 150, 180$ Hz in the quasi-steady state is shown in Fig. 4A. These results are obtained after a short transient, as the acoustic superlens rapidly reaches the quasi-steady state after a few cycles of the excitation. Strikingly, the resolution of the acoustic image drastically improves as virtual gain is added, reaching a maximum for $\gamma = \alpha \approx 180$ Hz. Similar to the simulations, we computed the corresponding spatial FT components in the $y$-direction and show the measured transmission normalized to the monochromatic excitation for each transverse wavenumber in Fig. 4B. Indeed, virtual gain stemming from complex frequency excitation significantly enhances the transmission of large transverse wave numbers associated with the evanescent spectrum of the image, explaining the observed resolution improvement. For $\gamma = \alpha \approx 180\,\text{Hz}$, large wavenumbers are transmitted with a six times enhancement compared to the harmonic excitation case. This effect is further corroborated in the case of a more complex 2D object consisting of a smiley face (Fig. 4C). The comparison between the normalized intensities at $\gamma = 0$ (harmonic) (top) and $\gamma = \alpha$ (bottom) shows a significant imaging resolution improvement, allowing to retrieve most of the object's subwavelength features.

*Discussion*

In this work, we have demonstrated that *virtual* gain obtained by precisely tailoring incident signals oscillating at complex frequencies can be used to compensate for the adverse effects of material loss in passive superlenses. We experimentally demonstrated this concept in a holey acoustic metamaterial superlens, verifying that, when the decay rate of the impinging excitation matches the intrinsic loss rate of the superlens, a complete restoration of the subwavelength features of the object is achievable, implementing a much improved superlens whose resolution is limited only by nonlocalities. Importantly, complex frequency excitations have a finite frequency bandwidth, which should be considered for the definition of the diffraction limit. Nevertheless, the requirement that the superlens reaches a quasi-steady state, which is necessary to be able to observe this phenomenon, implies that the structures support a high quality-factor, and as such the finite bandwidth of the excitation is narrow. Hence, the actual frequencies exciting the structure are still concentrated within a limited range around the central frequency, and our results clearly demonstrate deeply subwavelength imaging. Indeed, the demonstrated resolution enhancement is directly related to the virtual gain phenomenon, as further discussed in Supplementary Section S5. Our approach to compensate for loss and enhance the resolution of imaging devices does not require active elements, signal processing or feedback, and since the structure necessarily supports a large quality factor, it operates for a relatively long temporal range within the quasi-stationary regime, opening numerous opportunities for imaging, which may go beyond acoustics and be extended to nano-optics [3,4,37]. More broadly, our results pave the way for loss compensation in a variety of metamaterial, photonic and acoustic platform, offering a plethora of opportunities for imaging, sensing, computing and communications, and may be also extended to quantum technologies.

**Acknowledgments**

The work was supported in part by the Simons Foundation and the Air Force Office of Scientific Research.

**Figures**

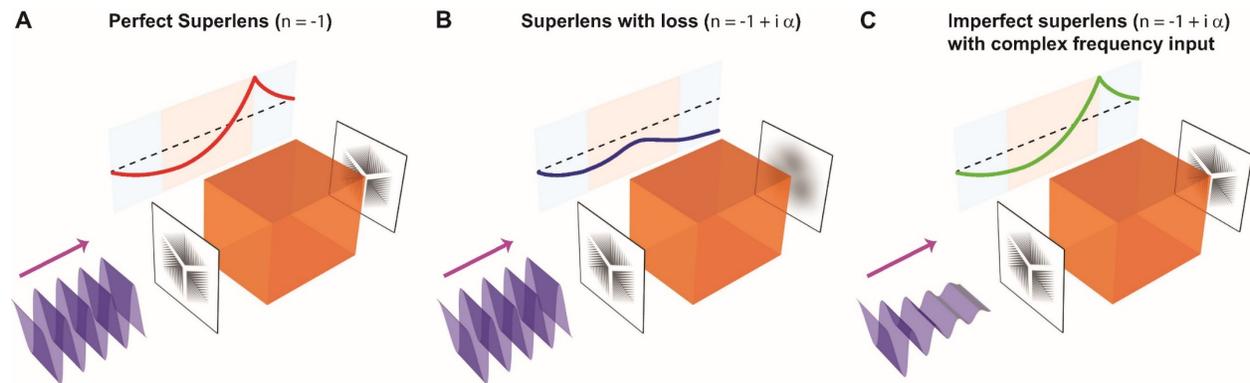

**Figure 1. Superlens response limited by material loss** (**A**) A superlens without loss amplifies evanescent fields, providing perfect imaging of an object at the image plane. (**B**) Material loss in the superlens is detrimental to the amplification of evanescent waves, resulting in limited resolution. (**C**) Excitation at a suitable complex frequency imparts virtual gain to a passive lossy superlens, enabling an enhancement in resolution and compensation of the detrimental effect of material loss.

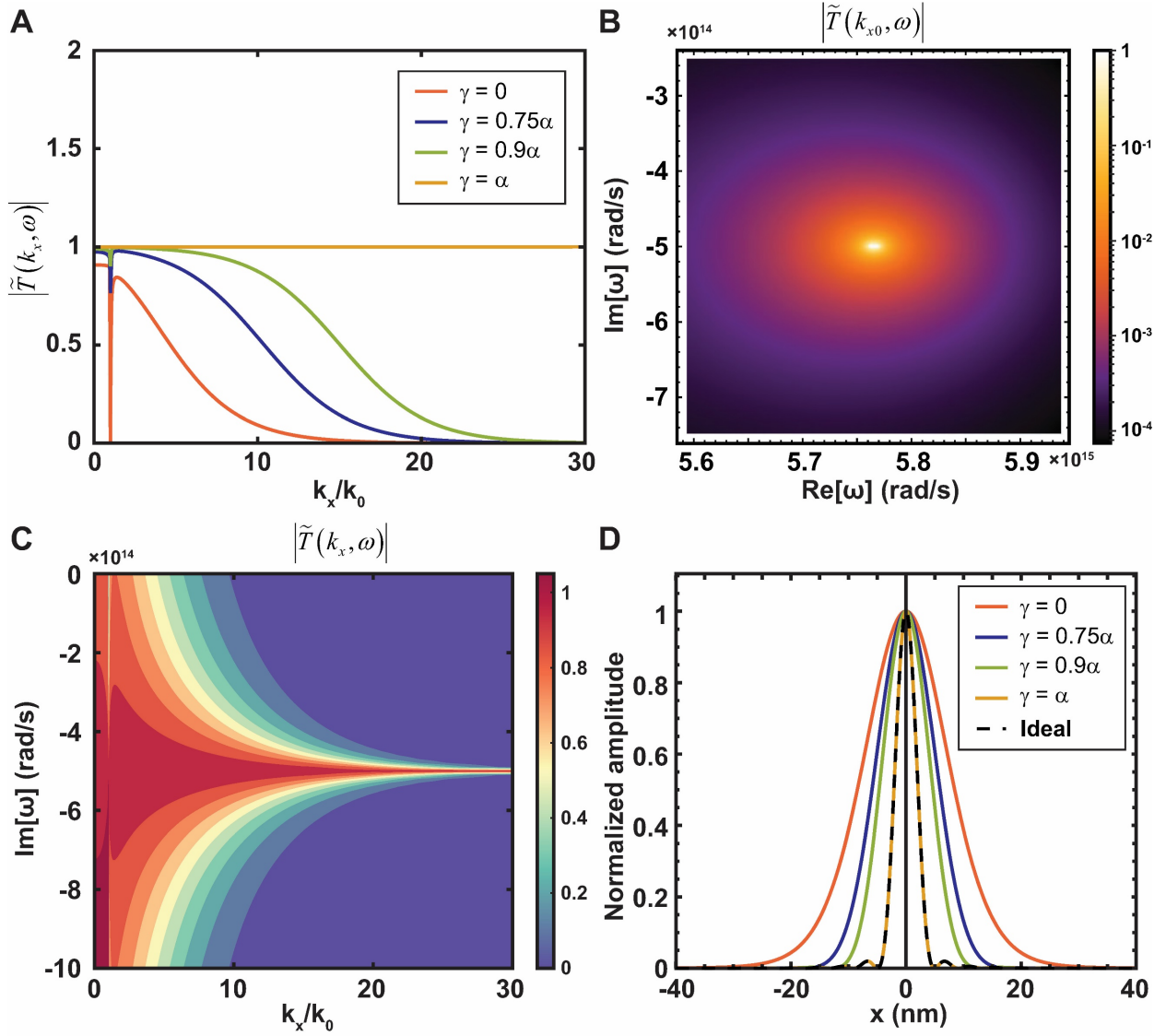

**Figure 2. Virtual gain compensating material absorption in a Drude superlens**. (**A**) Transmission versus transverse wavenumber for various decay rates $\gamma = 0$ (harmonic), $0.75\alpha$, $0.9\alpha$ and $\alpha$. The impinging waves oscillate at the same real part of $\omega_c$ with different decay rates. At $\gamma = \alpha$ (yellow curve), the dispersion becomes identical to the dispersion of the ideal. (**B**) Normalized transmission $|\tilde{T}(k_{x0},\omega)|$ for $k_{x0} = 30k_0$. The transmission is unitary at complex frequency $\omega_c = 5.77\times10^{15} - i\,5\times10^{14}$ rad/s. (**C**) Density map of the transmission as a function of decay rate and $k_x$, confirming that transmission is unity when the decay rate is equal to the intrinsic loss rate of the system. (**D**) Point spread function in the image plane for $\gamma = 0$, $0.75\alpha$, $0.9\alpha$ and $\alpha$. As the decay rate approaches the loss rate, the function becomes close to the lossless scenario (dotted curve). The width of the function at $\gamma = \alpha$ is about five times smaller than the width of the function at $\gamma = 0$.

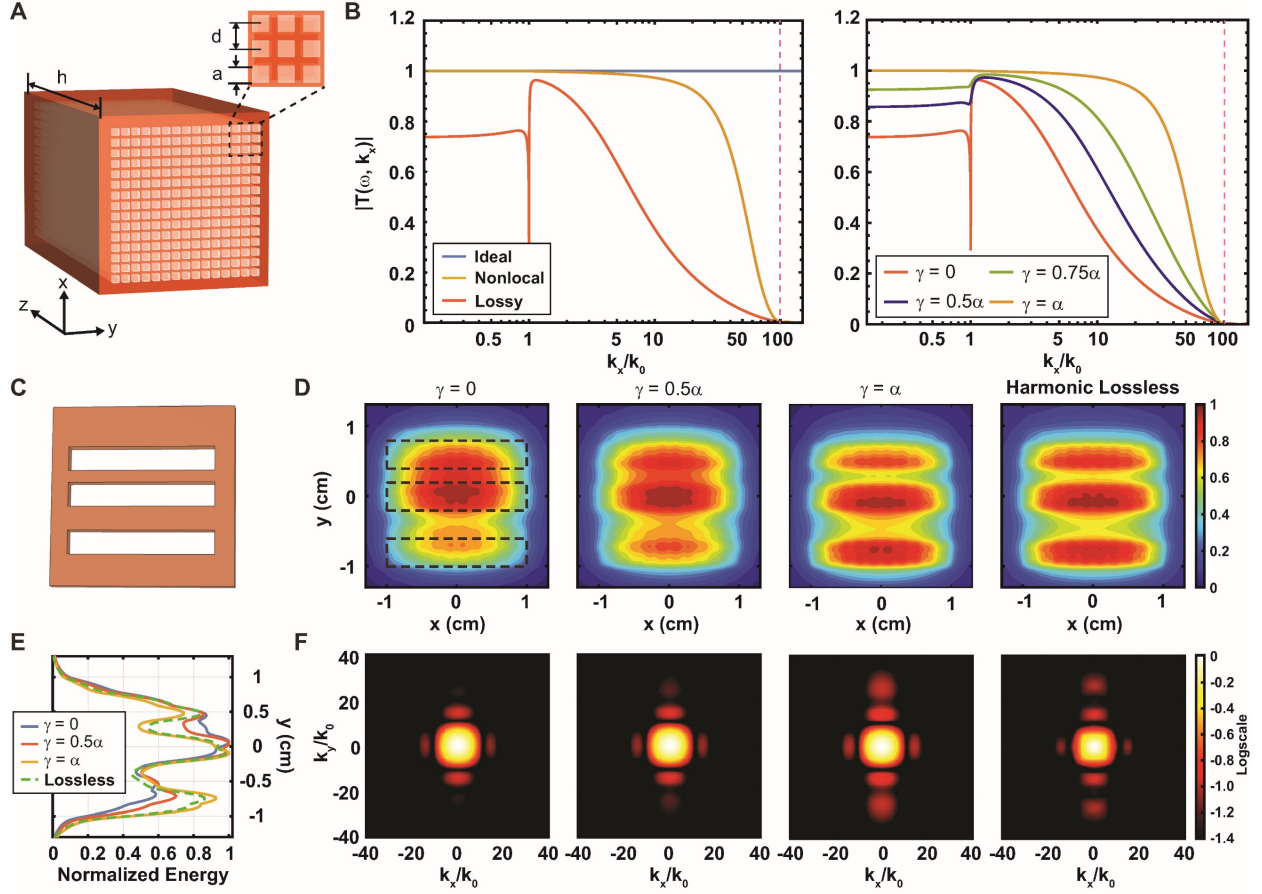

**Figure 3. Holey acoustic metamaterial under complex frequency excitation.** (**A**) Schematic of the acoustic metamaterial with thickness $h = 10$ cm. The size of the square holes is $a = 0.15$ cm, and the lattice constant is $d = 0.2$ cm. (**B-left**) Transmission versus transverse wavenumber. The transmission at the first FP resonance is not unity for large $k_x$ (yellow) due to nonlocality, such that its response is fundamentally limited by $2\pi/d \approx 100 k_0$ (red dotted line). In the lossy case, the dispersion (red) deviates from the lossless case (yellow). (**B-right**) Incident waves impinge on the structure at the first FP resonant frequency with different complex frequencies. For $\gamma = \alpha$, the dispersion is identical to the dispersion of the lossless scenario (yellow in the left panel). (**C**) Schematic of the object plate with three lines. The width is 0.4 cm, and the gaps are 0.2 and 0.4 cm. (**D**) Sound intensities for various decay rates $\gamma = 0$, $0.5\alpha$ and $\alpha$. The dotted boxes represent the original three-line object. The image at $\gamma = \alpha$ becomes consistent with the image in the harmonic lossless case. (**E**) Image cut of sound intensities for each case. (**F**) 2D spatial Fourier components of the sound maps in Fig. 3D (log scale).

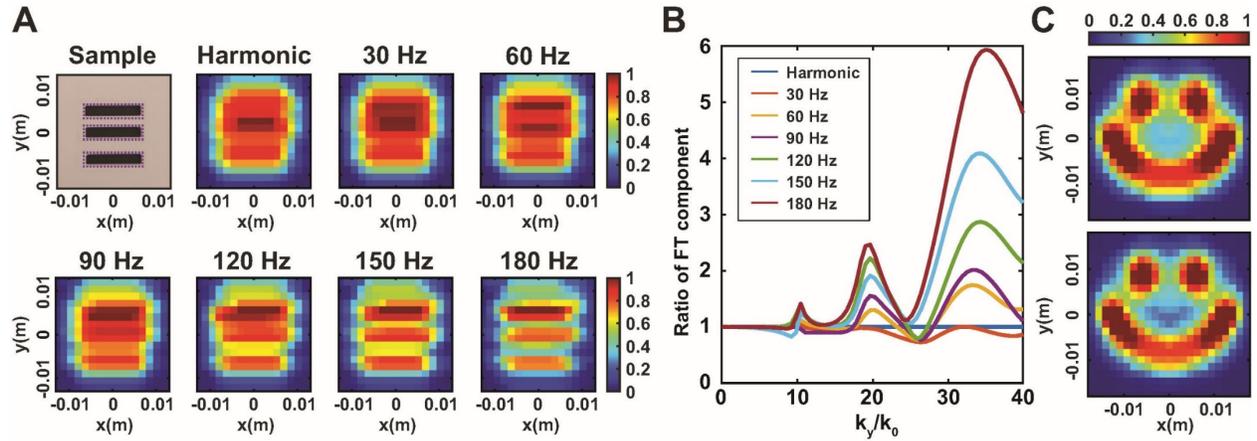

**Figure 4. Experimental demonstration of virtual acoustic superlens.** (**A**) Three-line object and sound maps for each case. The maps represent acoustic energy normalized by each maximum. Once the decay rate is close to the loss rate $\alpha \approx 180\,\text{Hz}$, the image map is optimally resolved. (**B**) 1D Fourier components for different decay rates normalized to the same transverse wavenumber in the harmonic excitation case. (**C**) Comparison of smiley face images under harmonic (top) and complex frequency (bottom) excitations. Here the decay rate of the complex excitation is 180 Hz.